\documentclass[letter]{jpsj3} 
\title{Momentum Dependent Local-Ansatz Wavefunction from Weak to Strong
Electron Correlations}

\author{
M. Atiqur R. \textsc{Patoary}\thanks{E-mail address:
k108609@eve.u-ryukyu.ac.jp}, Sumal \textsc{Chandra},  and  Yoshiro \textsc{Kakehashi}\thanks{yok@sci.u-ryukyu.ac.jp}
}

\inst{
Department of Physics and Earth Sciences,
Faculty of Science, \\
University of the Ryukyus, \\
1 Senbaru, Nishihara, Okinawa, 903-0213, Japan
}

\abst{
\begin{center}
( To be published in Journal of the Physical Society of Japan )
\end{center}

Momentum dependent local-ansatz (MLA) wavefunction describes accurately electron
correlations from the weak to intermediate Coulomb interaction regimes.   
We point out that the MLA can describe the correlations from the weak to strong Coulomb 
interaction regimes  by modifying the starting wavefunction from the Hartree-Fock (HF)
type to an alloy-analogy (AA) type wavefunction. Numerical results based on  
the half-filled band Hubbard model on the hypercubic lattice in infinite 
dimensions show up that the new wavefunction yields  the ground-state 
energy  lower than the Gutzwiller wavefunction (GW) in  the 
whole Coulomb interaction regime. Calculated double occupation number is 
smaller than the result of the GW in the metallic regime, and is finite 
in the insulator regime. Furthermore, the momentum distribution shows a distinct 
momentum-dependence in both the metallic and insulator regions, 
which are qualitatively different from those of the GW.}

\kword{variational method, electron correlations, Gutzwiller
wavefunction, local ansatz, Hubbard model, metal-insulator transition, critical
Coulomb interaction, infinite dimensions}

\begin{document}
\maketitle

%
%
%
The variational theory is a useful tool for studying 
the ground-state properties of correlated electron system such as the magnetism,
the heavy-electron behavior, the metal-insulator transition, and the high-temperature
superconductivity~\cite{gutz63,gutz64,gutz65,stoll77,stoll78,stoll80,fulde95,pfaze90,
dbaer87,mzie97,dbae00,bahe10,dtahara08,hyoka11,tmisawa11,tschi12}. 
There the form of the wavefunction is crucial to describe electron correlations.
The  Gutzwiller  wavefunction (GW) is one of the popular 
wavefunctions, because of its simplicity and applicability to realistic systems~\cite{gutz63,gutz64,gutz65}. 
It describes electron correlations by making use of a projection operator 
$\Pi_{i} (1-g\hat n_{i\uparrow}\hat n_{i\downarrow})$ onto the Hartree-Fock (HF) 
wavefunction $|\phi_0\rangle$;  
$|\Psi_{\rm GW}\rangle =\big[\prod_{i}(1-g \hat n_{i\uparrow} {\hat n_{i\downarrow}})\big]|\phi_0\rangle $.
Here $\hat n_{i\sigma}$ is the number operator for an electron on site $i$ with 
spin $\sigma$, variational parameter $g$ reduces the amplitudes 
of doubly occupied states on  local orbitals.  The local-ansatz approach (LA) proposed 
by Stollhoff and Fulde is an alternative method which is simpler 
than the GW~\cite{stoll77,stoll78,stoll80,fulde95}.  The LA wavefunction 
takes into account the states expanded by the residual Coulomb interactions
$\{ O_{i} \}=\{ \delta \hat n_{i\uparrow}\delta \hat n_{i\downarrow} \}$;
$|\Psi_{\rm LA}\rangle = \big[ \prod_{i} 
(1 - \eta^{}_{\rm \, LA} O_{i}) \big]|\phi_{0} \rangle$. Here $\delta \hat n_{i\sigma}=
\hat n_{i\sigma} - \langle \hat n_{i\sigma} \rangle_{0}$, $\langle \hat n_{i\sigma}
\rangle_{0}$ being the HF average of electron number on site $i$ with spin
$\sigma$. The  amplitude $\eta^{}_{\rm \, LA}$ is determined variationally. 

Although the GW and the LA have been applied to a number of correlated electron systems, they are 
not sufficient for the description of correlations from the weak to strong interaction regimes.
In fact, both the GW and the LA do not describe exactly the weakly 
correlated regime due to the use of limited Hilbert-space. Furthermore in 
the strong Coulomb interaction regime, the GW yields the Brinkman-Rice atom ($i. e.$, no charge 
fluctuation on an atom) instead of the insulator solid in infinite 
dimensions ($d=\infty$)~\cite{br70}. In order to overcome the difficulty 
in the weak Coulomb interaction regime and to improve the behaviors in 
the intermediate Coulomb interaction regime, we have recently proposed the 
momentum-dependent local ansatz wavefunction (MLA)~\cite{kakeh08,pat11}. 
The MLA wavefunction controls the amplitudes of momentum-dependent two 
particle states to be best, and  much improves  the  Fermi liquid properties 
of the GW and the LA in those regimes.

The wavefunction which accurately describes the strong Coulomb interaction regime has been proposed by 
Baeriswyl~\cite{dbaer87, mzie97,dbae00,bahe10}. The Baeriswyl wavefunction (BW) 
is constructed by applying a hopping operator $\hat T$ onto the atomic  
wavefunction $|\Psi_{\infty}\rangle$; 
$|\Psi_{\rm BW}\rangle = e^{-\eta\hat T}|\Psi_{\infty}\rangle$. Here 
$\hat T=-\sum_{i,j,\sigma} t_{ij} {a}^\dag_{i\sigma} {a}_{j\sigma} $ is the
kinetic energy operator. $t_{ij}$ denotes the transfer integral between 
sites $i$ and $j$, $ a_{i\sigma}^{\dagger}$ ($ a_{i\sigma}$) being the creation 
(annihilation) operator for an electron on site $i$ with spin $\sigma$.  
The operator $e^{-\eta\hat T}$ with a variational parameter $\eta$ 
describes electron hopping from the atomic state and 
suppresses the configurations with high kinetic energy. 
Although the BW describes well the insulator state in the strong correlation 
regime,  it is not easy to describe the metallic state from this viewpoint. 
There is no other variational wavefunction which is suitable for both  the weak 
and strong Coulomb  interaction regimes and  can be analytically treated, as far as we know. 

In this letter, we point out that a modified  MLA wavefunction which  starts
from the alloy-analogy (AA) wavefunction instead of the HF one describes the strongly 
correlated regime reasonably so that one can go beyond the GW in both 
the  weak and  strong Coulomb interaction regimes. The wavefunction provides us with a new 
tool to describe various systems with intermediate Coulomb interaction strength 
such as the cuprates and the iron pnictides superconductors.

We adopt in this letter the single-band Hubbard model as follows~\cite{gutz63,hub-I, hub-II}:
\begin{eqnarray}
H=\sum_{i\sigma}(\epsilon_0-\mu)\hat n_{i\sigma}
+ \sum_{ij\sigma} t_{ij}  a_{i\sigma}^\dagger a_{j\sigma} 
+U \sum_i \hat n_{i\uparrow} \hat n_{i\downarrow}.
\end{eqnarray} 
Here $\epsilon_{0}$ ($\mu $) is the atomic level (chemical potential), 
$t_{ij}$ is the transfer integral between  sites $i$ and $j$.  
$U$ is the intra-atomic Coulomb energy parameter.  

The construction of the MLA wavefunction is rather simple. We expand  first 
the LA wavefunction up to the first order with respect to the Coulomb interaction 
strength $U$ and observe that each coefficient of the two-particle excited states
in the first-order wavefunction is momentum independent in contradiction 
to the exact result. Therefore, we introduce 
a new set of local operators $\{\tilde O_i\}$ with the momentum-dependent 
variational parameters $\eta_{k'_2 k_2 k'_1 k_1}$ which yields the correct weak 
Coulomb interaction limit:
\begin{eqnarray}
\tilde{O_i} = \sum_{k_1 k'_1 k_2 k'_2} 
\langle k'_1 | i\rangle \langle i | k_1 \rangle 
\langle k'_2 | i\rangle \langle i | k_2 \rangle 
\ \eta_{k'_2 k_2 k'_1 k_1} 
\delta ({a_{k'_2 \downarrow}^\dagger} a_{k_2 \downarrow }) 
\delta ({a_{k'_1 \uparrow }^\dagger} a_{k_1 \uparrow }) \ . 
\end{eqnarray} 
Here $\langle i|k \rangle = \exp (-i\boldsymbol{k}\cdot \boldsymbol{R}_{i}) / \sqrt{N}$ 
is an overlap integral between the localized orbital and the Bloch state with
momentum $\boldsymbol{k}$, $\boldsymbol{R}_{i}$ denotes atomic
position, and $N$ is the number of sites.
$ a_{k \sigma}^{\dagger}$ ($ a_{k \sigma}$) denotes a creation 
(annihilation) operator for an electron with momentum $\boldsymbol{k}$ 
and spin $\sigma$, and 
$\delta(a^{\dagger}_{k^{\prime}\sigma}a_{k\sigma})= 
a^{\dagger}_{k^{\prime}\sigma}a_{k\sigma} - \langle
a^{\dagger}_{k^{\prime}\sigma}a_{k\sigma} \rangle_{0}$.
We then construct the MLA wavefunction  as follows~\cite{kakeh08,pat11}.
\begin {equation}
 | \Psi_{\rm MLA} \rangle = \prod_i (1-\tilde{O_i}) | \phi_0 \rangle.
\label{wmlahf}
\end{equation}  
The best wavefunction is chosen by controlling the variational parameters
in the momentum space. Hereafter we refer the wavefunction (\ref{wmlahf}) to the MLA-HF wavefunction. 

The MLA-HF wavefunction does suitably describe the electron correlations in the 
weak and intermediate Coulomb interaction, but it cannot suppress loss of 
Coulomb repulsion in the strong interaction regime. To improve the difficulty
we propose here to change the starting wavefunction from the HF wavefunction 
to the alloy-analogy (AA) wavefunction which is suitable in the strong Coulomb 
interaction regime. 

The concept of the AA can be traced back to Hubbard's original work 
on electron correlations~\cite{hub-III}. In the strong Coulomb interaction regime,
electrons with spin $\sigma$ move slowly from site to site due to electron correlations,
and therefore should feel a potential $U$ instead of the HF average
potential $U\langle \hat n_{i-\sigma} \rangle_0$,
when the  opposite spin electron is on the same site. Hubbard regarded this system as 
an alloy with different random potentials $\epsilon_0+U$ and $\epsilon_0$. 
The AA Hamiltonian is then defined by  
\begin{eqnarray}
H_{\rm AA} =\sum_{i\sigma}(\epsilon_0-\mu+U n_{i-\sigma})\hat n_{i\sigma}
+ \sum_{ij\sigma} t_{ij}  a_{i\sigma}^\dagger  a_{j\sigma} 
-U \sum_i n_{i\uparrow} n_{i\downarrow}.
\label{aa}
\end{eqnarray} 
Since  the motion of electrons with opposite spin are  treated  
to be static in the AA approximation, related operators $\{\hat n_{i\sigma} \}$ are regarded as a random 
static  $C$ number $n_{i\sigma}$ ($0$ or $1$). Each configuration  $\{ n_{i\sigma} \}$ is considered as a 
snapshot in time development.

We adopt the AA ground-state  wavefunction $\phi_{\rm AA}$ for the Hamiltonian $H_{\rm AA}$, 
and propose a new ansatz, which we call the MLA-AA wavefunction,  as follows. 
\begin {equation}
 | \Psi_{\rm MLA-AA} \rangle = \prod_i (1-\tilde{O_i}) | \phi_{\rm AA} \rangle .
\end{equation}  
The local operators $\{\tilde O_i\}$ with  variational parameters
$\eta_{\kappa'_2 \kappa_2 \kappa'_1 \kappa_1}$ have been modified as  
\begin{eqnarray}
\tilde{O_i} = \sum_{\kappa'_2 \kappa_2 \kappa'_1 \kappa_1} 
\langle \kappa'_1 | i\rangle \langle i | \kappa_1 \rangle 
\langle \kappa'_2 | i\rangle \langle i | \kappa_2 \rangle 
\ \eta_{\kappa'_2 \kappa_2 \kappa'_1 \kappa_1} 
\delta ({a_{\kappa'_2 \downarrow}^\dagger} a_{\kappa_2 \downarrow }) 
\delta ({a_{\kappa'_1 \uparrow }^\dagger} a_{\kappa_1 \uparrow }) \ . 
\end{eqnarray} 
Here $ a_{\kappa \sigma}^{\dagger}$ and $ a_{\kappa \sigma}$ are the creation 
and annihilation operators which diagonalize the Hamiltonian $H_{\rm AA}$, and 
$\delta(a^{\dagger}_{\kappa^{\prime}\sigma}a_{\kappa\sigma})= 
a^{\dagger}_{\kappa^{\prime}\sigma}a_{\kappa\sigma} - \langle
a^{\dagger}_{\kappa^{\prime}\sigma}a_{\kappa\sigma} \rangle_{0}$.
It should be noted that the MLA-AA wavefunction reduces to the MLA-HF by replacing 
the random potential $U n_{i-\sigma}$  with the HF one, $i.e.$, 
$U\langle\phi_0| \hat n_{i-\sigma}|\phi_0 \rangle$, so that $\Psi_{\rm MLA-AA}$ and 
$\Psi_{\rm MLA-HF}$ are mutually connected to each other via a suitable parameter
which interpolates between the two wavefunctions.

We can obtain  the ground-state energy for the MLA-AA wavefunction
within the single-site approximation (SSA) taking the same steps as in the 
MLA-HF~\cite{kakeh08,pat11}. The correlation energy per atom is then given by
\begin{equation}
{\epsilon_c} =\frac{  - \langle {\tilde{O_i}}^\dagger\tilde{H} \rangle_0 
- \langle \tilde{H}  \tilde{O_i}\rangle_0 
+ \langle {\tilde{O_i}}^\dagger \tilde{H} \tilde{O_i} \rangle_0 }
{1+ \langle {\tilde{O_i}}^\dagger  \tilde{O_i} \rangle_0} \ .
\label{ec7}
\end{equation}
Here $\tilde H=H-\langle H\rangle_0 $. Note that the average $\langle \sim \rangle_0$ is now 
taken with respect to the AA wavefunction. The total energy per atom should be  obtained 
by taking the configurational average; $\langle H \rangle=\overline{ \langle H \rangle_0}+N \overline{\epsilon_c}$.
Here the upper bar denotes the configurational average.
Each term in the correlation energy can be calculated 
by making use of Wick's theorem.

We obtain a self-consistent equation with variational 
parameters using the minimum energy condition. 
\begin{eqnarray}
(\Delta E_{{\kappa^{\prime}_{2}\kappa_{2}\kappa^{\prime}_{1}\kappa_{1}}} - \epsilon_{\rm c})
\eta_{\kappa^{\prime}_{2}\kappa_{2}\kappa^{\prime}_{1}\kappa_{1}}  \hspace{5mm} \nonumber \\
& &  \hspace{-55mm}
+ \dfrac{U}{N^{2}}
\Big[
\sum_{k_{3}k_{4}}
f(\tilde{\epsilon}_{\kappa_{3}\uparrow})f(\tilde{\epsilon}_{\kappa_{4}\downarrow})
\eta_{\kappa^{\prime}_{2}\kappa_{4}\kappa^{\prime}_{1}\kappa_{3}}
- \sum_{\kappa_{3}\kappa^{\prime}_{4}}
f(\tilde{\epsilon}_{\kappa_{3}\uparrow})
(1-f(\tilde{\epsilon}_{\kappa^{\prime}_{4}\downarrow}))
\eta_{\kappa^{\prime}_{4}\kappa_{2}\kappa^{\prime}_{1}\kappa_{3}}  \nonumber \\
& & \hspace*{-55mm}
- \sum_{\kappa^{\prime}_{3}\kappa_{4}}
(1 - f(\tilde{\epsilon}_{\kappa^{\prime}_{3}\uparrow}))
f(\tilde{\epsilon}_{\kappa_{4}\downarrow})
\eta_{\kappa^{\prime}_{2}\kappa_{4}\kappa^{\prime}_{3}\kappa_{1}}
+ \sum_{\kappa^{\prime}_{3}\kappa^{\prime}_{4}}
(1 - f(\tilde{\epsilon}_{\kappa^{\prime}_{3}\uparrow}))
(1 - f(\tilde{\epsilon}_{\kappa^{\prime}_{4}\downarrow}))
\eta_{\kappa^{\prime}_{4}\kappa_{2}\kappa^{\prime}_{3}k_{1}}
\Big] = U \, . \ \ \ \
\label{sceq}
\end{eqnarray}
Here $\Delta E_{\kappa^{\prime}_{2}\kappa_{2}\kappa^{\prime}_{1}\kappa_{1}} $ is 
the  two-particle excitation energy  given by 
$ \Delta E_{\kappa^{\prime}_{2}\kappa_{2}\kappa^{\prime}_{1}\kappa_{1}} 
= \epsilon_{\kappa^{\prime}_{2}\downarrow} - \epsilon_{\kappa_{2}\downarrow}
+ \epsilon_{\kappa^{\prime}_{1}\uparrow} - \epsilon_{\kappa_{1}\uparrow}$.
$\epsilon_{\kappa \sigma}$ denotes one-electron eigen value energy for $H_{\rm AA}$,
and  $\tilde\epsilon_{\kappa \sigma}=\epsilon_{\kappa \sigma}-\mu$. $f(\epsilon)$
is the Fermi distribution function at zero temperature. To solve  the 
equation approximately, we make use of an interpolate solution which is valid 
in both the weak Coulomb interaction limit and the atomic limit~\cite{kakeh08}.
\begin{eqnarray}
\eta_{\kappa^{\prime}_{2}\kappa_{2}\kappa^{\prime}_{1}\kappa_{1}}(\tilde\eta,\epsilon_c) = 
\frac{U \tilde\eta}
{\Delta E_{{\kappa^{\prime}_{2}\kappa_{2}\kappa^{\prime}_{1}\kappa_{1}}} - \epsilon_c} \ .
\label{etaint}
\end{eqnarray}
Here $\tilde\eta = [1 - \eta (1 - 2 \langle n_{i\uparrow} \rangle_{0}) 
(1 - 2 \langle n_{i\downarrow} \rangle_{0})]$.
The best values of $\tilde \eta$ and $\epsilon_c$ are determined again variationally.

In the SSA, the correlation energy $\epsilon_c$ can be expressed by the 
local density of states $\rho_{i\sigma}(\epsilon)$ for the AA Hamiltonian (\ref{aa}).
Because there is no translational symmetry  due to random potential, we calculated
$\rho_{i\sigma}(\epsilon)$ by means of the Coherent Potential Approximation 
(CPA)~\cite{shiba71,ehr76}.

To examine the validity of the MLA-AA as well as MLA-HF, we have performed the  
numerical calculations for the half-filled band Hubbard model with nearest neighbour 
transfer integral on the  hypercubic 
lattice in infinite dimensions, where the SSA works best~\cite{metz89,kakeh04}.
We assumed here the non-magnetic case.  In this 
case, the density of states for non-interacting system is given 
by $\rho(\epsilon) = (1/\sqrt{\pi}) \exp (-\epsilon^{2})$~\cite{metz89}. 
The energy unit is chosen to be $\int d\epsilon \rho (\epsilon)\epsilon^{2} = 1/2$.
The characteristic band width $W$ is given by $W=2$ in this unit.

\begin{figure}
\includegraphics[scale=0.5,angle=-90]{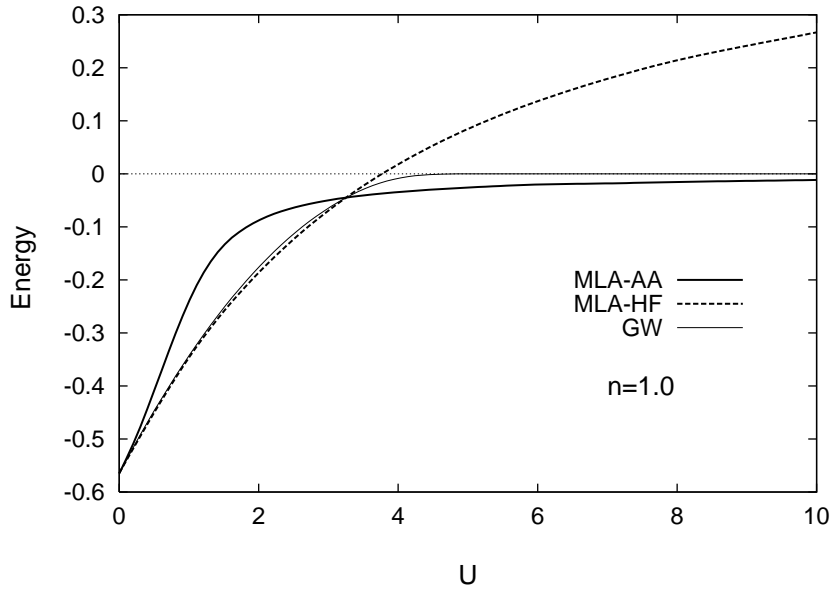}
\caption{\label{energy} The energy vs Coulomb interaction energy
$U$ curves in the MLA-AA (solid curve), MLA-HF (dashed curve),  and 
GW (thin solid curve) for the electron number $n=1.0$.}
\end{figure}

Figure 1 shows the results of the ground-state  energy vs Coulomb 
interaction energy curves. In the weak Coulomb interaction regime 
($U/W\lesssim 1$), the  total energy of the MLA-HF  is lower than the GW. On the other hand, 
we observe that the MLA-AA gives lower energy in comparison with the GW in the  
strong Coulomb interaction regime ($U/W>1$). We obtain the critical Coulomb 
interaction $U_{c2}=3.40$ at which the effective mass diverges. But before $U$ 
approaches $U_{c2}$  we find that the AA state showing the insulating state is stabilized, and
the metal-insulator transition occurs at the critical Coulomb interaction 
$U_c=3.26$. The transition is the first order in the present approach, and is consistent with 
the result of numerical renormalization group method (NRG), although the calculated $U_{c2}$ 
is somewhat smaller that obtained by NRG ( $i.e.$, $U_{c2}=4.1$ ).\cite{rbulla99}.   The MLA-HF leads to the total energy lower 
than that of the GW up to the critical Coulomb interaction $U_{c}$ and  the 
MLA-AA gives the same behavior in the range $U>U_{c}$. More important is 
that the MLA scheme gives lower energy for overall Coulomb  interaction regime 
and therefore can overcomes the limitation of the GW.

\begin{figure}
\includegraphics[scale=0.5, angle=-90]{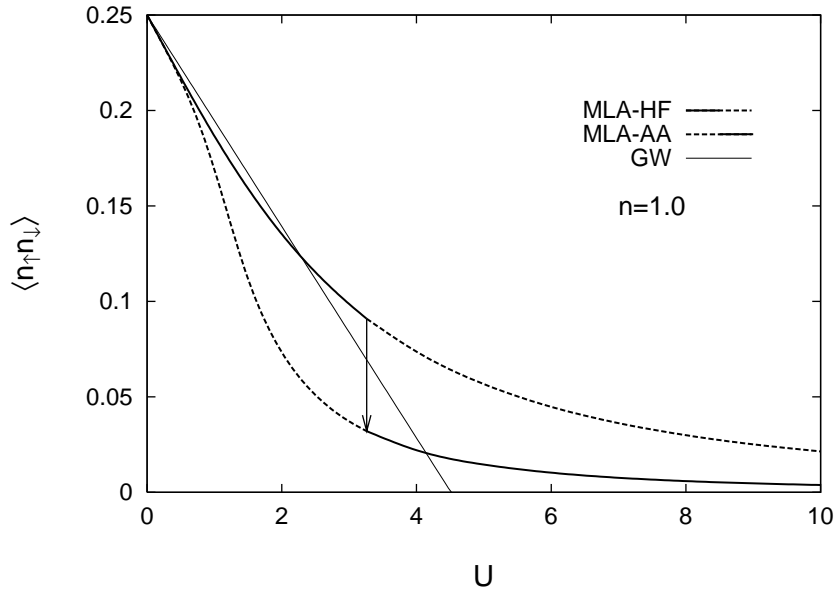}%
\caption{\label{dbl}The double occupation number  $\langle n_{\uparrow}n_{\downarrow} \rangle$ 
vs Coulomb interaction energy $U$ curve for the  electron number $n=1.0$ in the  MLA-HF (solid-dashed curve),
MLA-AA (dashed-solid curve), and GW (thin solid curve). The arrow shows a jump from the metallic state to the
insulator at $U_c=3.26.$}
\end{figure}
%

We present in Fig. 2 the double occupation number
$\langle n_{\uparrow}n_{\downarrow} \rangle$ as a function of  Coulomb 
interaction energy $U$ at half-filling.  It decreases  
from $1/4$ with increasing Coulomb interaction, so as  to reduce the 
loss of Coulomb energy $U$.
The MLA-HF state  reduces more the double occupancy as compared with 
that of the GW in the weakly correlated region, and jumps to 
the MLA-AA state at $U_{c}$. In the strongly correlated regime,
the MLA-AA gives finite value of double occupancy, while the GW gives 
the Brinkman-Rice atom.  This verifies the improvement  of the Brinkman-Rice atom.
The double occupancy for the MLA scheme  at $U_c$ is $0.032$, and is consistent 
with the result of the Quantum Monte Carlo ($\approx 0.024$) \cite{nblumer02}, though 
the latter uses the semi-elliptical density of states.

\begin{figure}
\includegraphics[scale=0.5, angle=-90]{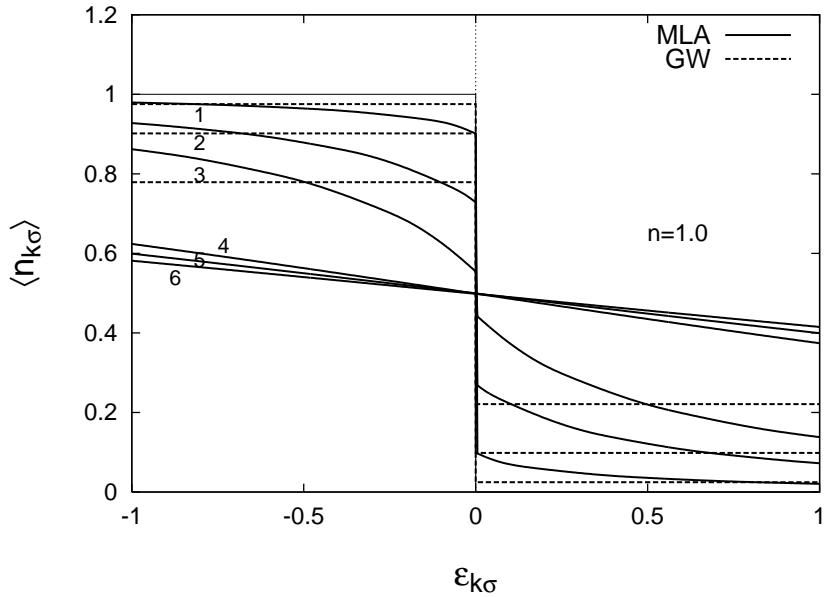}%
\caption{\label{momentum} The momentum distribution as a function of
energy  $\epsilon_{k\sigma}$ for various Coulomb interaction energy parameters
$U=1.0,2.0,3.0,4.0,5.0,6.0$ at half-filling. The MLA: solid curve, 
the GW: dashed curve, and the HF: thin solid curve.}
\end{figure}

The momentum distribution for the MLA shows a clear 
momentum-dependence as a function of the HF one electron energy 
$\epsilon_{k\sigma}$ as shown in Fig. 3. It decreases monotonically 
with increasing $\epsilon_{k\sigma}$ and shows a jump at the Fermi energy.
On the other hand, the distribution for the GW is constant below and above the Fermi 
level~\cite{gutz63,gutz64,gutz65}. The jump decreases with increasing $U$, and 
disappears beyond $U_c$, indicating the insulating state. The curve becomes 
flatter with further increase of  $U$.

In summary, we have proposed a new momentum-dependent local ansatz wavefunction
(MLA) which allows us to describe electron correlations  starting from 
both the Hartree-Fock (HF) and the alloy-analogy (AA) limits. The former ($i.e.$, MLA-HF)
describes the Fermi-liquid state, while the latter ($i.e.$, MLA-AA) describes the 
insulator state. We have performed the  numerical calculations for 
the half-filled band Hubbard model  on the  hypercubic lattice in infinite 
dimensions, and  demonstrated  that the ground state energy for the MLA is 
lower than the GW in the whole range of Coulomb interaction. The MLA yields the 
metal-insulator transition at $U_c=3.26$. The double occupation number is suppressed 
in the weak and intermediate Coulomb interaction regimes as compared with the 
GW,  jumps at $U_{c}$, and  remains finite in the strongly correlated regime as it 
should be.  Finally, we found  the momentum distribution functions showing  a distinct  
momentum dependence in both the metallic and insulator regimes. These results 
indicate that the MLA approach can overcome the limitations of the original MLA,
and goes beyond the GW in both the weak and strong $U$ regimes.

The present work is supported by a Grant-in-Aid for
Scientific Research (22540395).

\end{document}